\documentclass[doublespace,preprint,fleqn,natbib]{svjour2}

\bibpunct{[}{]}{;}{n}{}{,} 
\smartqed  
\usepackage{graphicx}
\usepackage[T1]{fontenc}
\usepackage{bm}
\usepackage{amsmath}
\journalname{Granular Matter} 

\begin{document}

\title{A discrete element study of settlement in vibrated granular layers: role of
contact loss and acceleration}

\author{A. Karrech  \and
        D. Duhamel  \and
        G. Bonnet   \and
        F. Chevoir  \and
        J.-N. Roux  \and
        J. Canou    \and
        J.-C. Dupla \and
}

\institute{Address: \at
           Université Paris-Est, Institut Navier, LAMI, Ecole des
           Ponts, 6 et 8 av Blaise Pascal, Cité Descartes, Champs Sur Marne, 77455 Marne la Vallée
           Cedex2, France.
           \and
           Prof D. Duhamel (Corresponding Author): \at
           Tel.: +33 1 64 15 37 28  Fax:  +33 1 64 15 37 41 \email{duhamel@lami.enpc.fr}
           \and
           Prof G. Bonnet: \at
           Université Paris-Est, Laboratoire de Modélisation et
           Simulation Multi-Echelle, 5 bd Descartes, Champs Sur Marne, 77454 Marne la Vallée
           Cedex2, France.
}

\date{Received: \today}

\maketitle

\begin{abstract}

This paper deals with the vibration of granular materials due to
cyclic external excitation. It highlights the effect of the
acceleration on the settlement speed and proves the existence of a
relationship between settlement and loss of contacts in partially
confined granular materials under vibration. The numerical
simulations are carried out using the Molecular Dynamics method,
where the discrete elements consist of polygonal grains. The data
analyses are conducted based on multivariate autoregressive models
to describe the settlement and permanent contacts number with
respect to the number of loading cycles.

\end{abstract}

\section{Introduction}
\label{intro} Ballast materials of railway platforms exhibit complex
behavior under repeated loading. With the increase of vehicle speeds
as well as comfort and safety requirements, understanding the
dynamics of these materials is becoming a crucial issue. Unlike
highly agitated granular materials which can be described with the
kinetic theory, ballasted layers are generally subjected to dense
flow where the trajectories of grains are correlated and the
collisions cannot be considered as randomly distributed in terms of
positions and velocities. Therefore, such a statistical approach may
be inappropriate to predict the response of railway platforms.

In this research work, the Molecular Dynamics method is used to
simulate granular samples made of polygonal grains under vibration.
Since its introduction by Cundall and Strack \cite{Cundall}, this
discrete elements method has proved its viability in describing
several mechanisms such as granular materials transport
\cite{Transport}, mixing \cite{Mixing}, segregation \cite{SegMix},
compaction \cite{Rosatocompaction} etc. Coupled with the
experimental approaches, the MD is now recognized as a fundamental
tool to investigate the behavior of dense granular materials.
Recently, Lu and McDowell \cite{Lu_Mc} developed an approach based
on the MD to investigate the permanent displacements in granular
beds, made of irregular shaped grains, under a single loading cycle.
The complex geometry was produced using spherical grains assemblies.
Although efficient in terms of grain shapes, this technique neglects
the inertial effects of the overlapping regions. The discrete
element method was also adopted by Lobo-Guerrero and Vallejo
\cite{lobo_vall} on railway platforms to investigate the effect of
grain degradation on granular bed response under cyclic loading. The
model took into account the rupture of grains using a criterion
based on the loading modes and force intensities. Similar approaches
such as the molecular dynamics method was used to simulate the
vibration of confined granular material made of polygonal components
\cite{azema,saussine}. In this paper, the target is to relate the
permanent settlement in partially confined samples to the loss of
contacts using statistical analysis. In the second section, a brief
description of the simulation method is suggested. The settlement
mechanism which represents the residual displacement under sleeper
is described through field cases simulation. In the third section, a
causality analysis based on multivariate autoregressive models
describing the settlement and contact loss is conducted.

\section{Simulation method}
\begin{figure}[h!]
\begin{center}
\begin{tabular}{c c}
(a) & (b)\\
  \includegraphics[width=0.45\hsize]{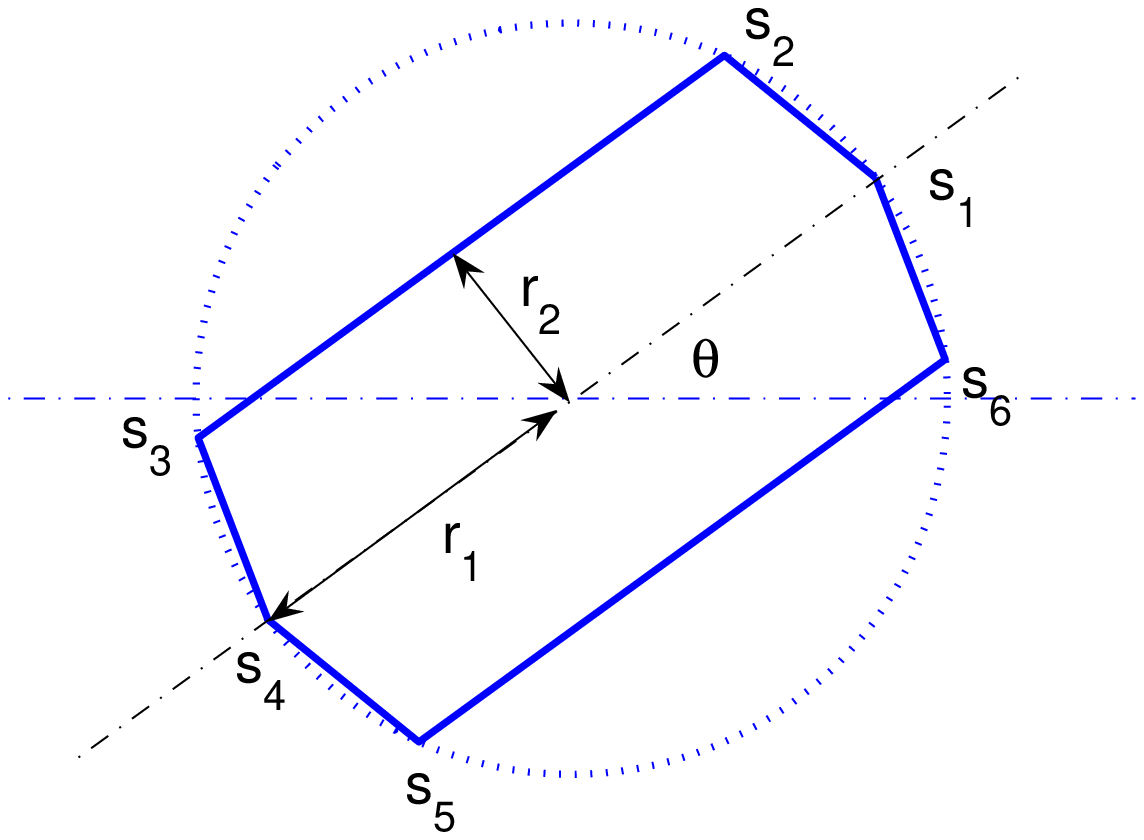} &
  \includegraphics[width=0.45\hsize]{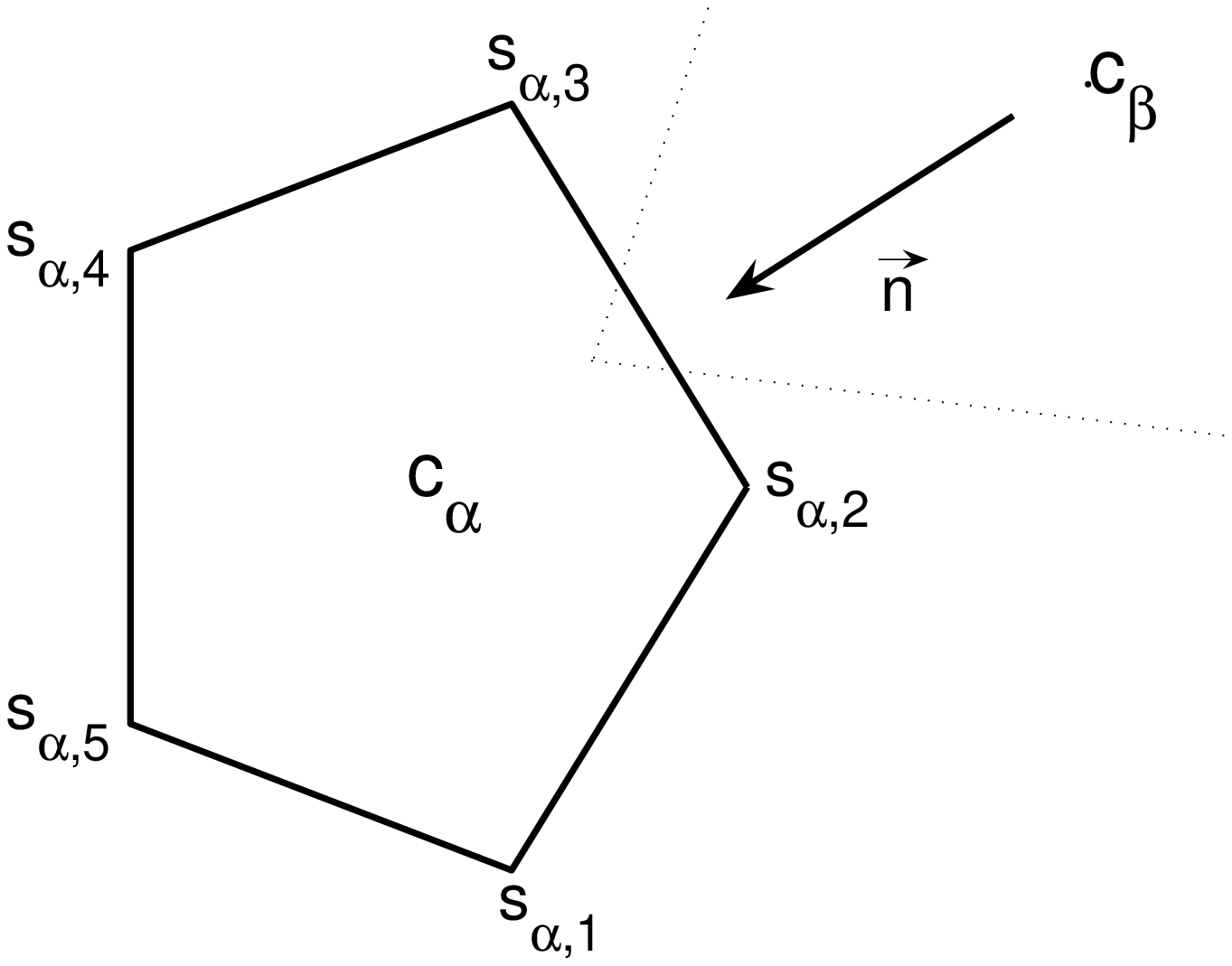}\\
\end{tabular}
\caption{\label{grains_shape} (a) Typical grain characterized by
$r_1$, $r_2$, and $\theta$, (b) Contact detecting using the
dichotomy method.}
\end{center}
\end{figure}

The sample is composed of polygonal grains which are described with
a set of ordered vertices $(s_{\alpha,i})_{i \in [|1,6|]}$. The
positions of the vertices can be evaluated using the distances $r_1$
and $r_2$ as well as the orientation $\theta$, which are shown in
figure \ref{grains_shape}-(a). These parameters follow uniform
bounded distributions: $ r_1 \sim U_{[r_{\min },r_{\max} ]}$, $r_2
\sim U_{[0.25 r_1 ,0.75r_1 ]}$, and $\theta \sim U_{[0;2\pi ]}$.

The interaction between grains takes place when an overlap is
detected (Figure \ref{grains_shape}-b). The dichotomy method is used
to calculate the shortest distance between the vertices. This leads
to the edges which are candidates for interaction. Among this list
of edges, the contact segment which relates the couple of
intersection points is obtained. It defines the tangential
component, $\boldsymbol{t}$, of the contact frame and its
perpendicular represents the normal component, $\boldsymbol{n}$. The
reference point of the frame, $o$, is defined by the middle of the
above mentioned intersection points. Once the contact frame, the
overlap, and the velocities are known, the contact forces acting on
a grain $\alpha$ can be written as follows:
\begin{equation}
  \label{eqa4}
    \begin{array}{l}
     \displaystyle F_n= F_n^e - \gamma_n v_n \\
     \displaystyle F_t=  \left\{%
                            \begin{array}{ll}
                                F_t^{e}  - \gamma_t v_t \qquad \qquad \quad \;\;   \mbox{if} \;  \| F_t\| \leq \mu F_n  \\
                                 sign(F_t^{e}) \mu F_n - \gamma_t v_t \quad  \mbox{if} \;\|F_t\| > \mu F_n \\
                            \end{array}%
                         \right. \\
  \end{array}
\end{equation}
The interaction forces acting on the particles contain elastic
terms; by assuming uniform pressure and shear stresses along the
contact surface, it can be shown that these elastic forces are
linear: $F_n^e = k_n u_x$ and $F_t^e = k_t u_y$, where $k_n=
\frac{E} {c_n(1 - \nu ^2 )}$ and $k_t= \frac{E} {c_t(1 - \nu ^2 )}$.
$E$ and $\nu$ are the materials properties (Young modulus and
Poisson's ratio) whereas the constants $c_n$ and $c_t$ can be
obtained experimentally using a compression test. The interactions
also enclose viscous terms denoted by the phenomenological constants
$\gamma_{n,t}$ which can be rewritten as dissipation fraction with
respect to the critical damping: $\gamma_{n,t} = \alpha_{n,t}
\sqrt{m k_{n,t}}$. Finally, the Coulomb friction is included through
the threshold $F_t \le \mu F_n$, where $\mu$ is the friction
coefficient. At the same time the moments are calculated using the
contact frame and the contact forces. The discrete element
simulations are carried out with the following parameters: density
of grains, $\rho_p = 2710 kg/m^3$, Young modulus, $E = 46.9 GPa$,
Poisson's ratio, $\nu = 0.25$, friction coefficient, $\mu = 0.8$,
viscous coefficients, $\alpha_n = 0.8$ and $\alpha_t = 0.1$, and
grain dimensions $r_{min}=r_{max}=5$ mm (unless differently
specified). It is worthwhile noticing that the interaction with the
wall is described the same way as the interactions between grains.

Using the contact forces between two grains $\alpha$ and $\beta$ at
the reference point $o$, it is possible to calculate the moments
around the centroid $c_{\alpha}$ using the identity $
\boldsymbol{M_{c_{\alpha}}} = \boldsymbol{M_o} + \boldsymbol{F_c}
\times \boldsymbol{oc_{\alpha}}$, where $\boldsymbol{F_c}$ is the
contact force applied on the grain $\alpha$ as expressed beforehand
in equation (\ref{eqa4}). The procedure is used for each component
of the sample. Once the moments and interactions are known, the
equations of motion can be integrated with respect to time using a
finite difference scheme \cite{Cundall, Karrech3, Karrech1},
according to the Molecular Dynamics method.

\subsection{Sample preparation, loading, and boundary conditions}
\begin{figure}[h!]
\begin{center}
\begin{tabular}{c}
  \includegraphics[width=1\hsize]{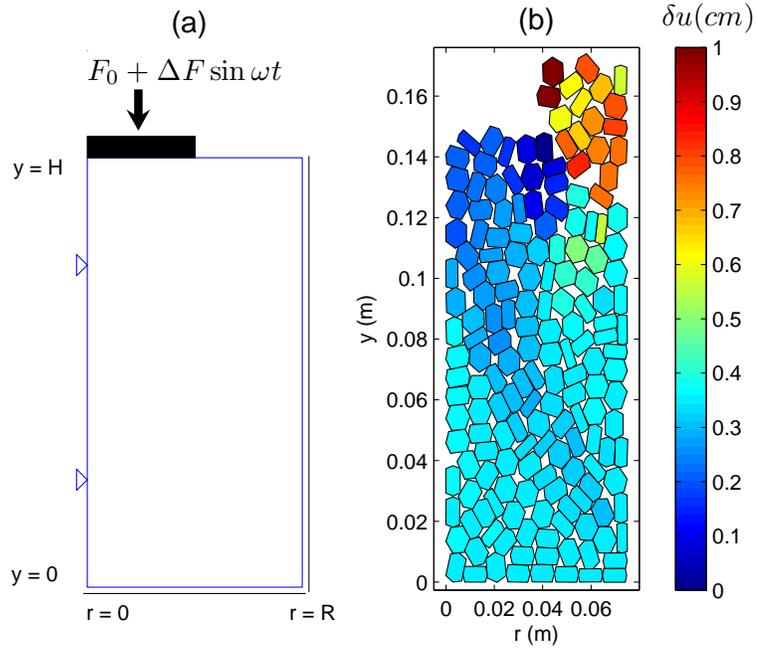}
\end{tabular}
\caption{\label{BC} (a) Boundary conditions and applied force (b)
Residual displacement of grain (the colorbar represents the
intensity of residual displacement in the y-direction in cm).}
\end{center}
\end{figure}

At the beginning of the process, the grains are subjected to the
gravity field until reaching the full equilibrium. The displacements
and rotations are calculated using the predictor-corrector algorithm
\cite{allen}, where the contact forces, the moments and the body
forces are taken into consideration. Once the equilibrium is
reached, the sample - of width $R = 75$ mm, height $H = 150$ mm and
number of grains 175 - is subjected to a sinusoïdal load of the form
$F=F_0 + \Delta F \sin \omega t$, where $\omega$ is the circular
frequency, $F_0$ is the initial force which is taken equal to
$0.5kN$, and $\Delta F$ is the force amplitude. The frequency and
force amplitude vary in such a way that the sleeper covers a wide
range of accelerations around g (gravity). The excitation is applied
on the inner half ($0 \le r \le 0.5R$) and upper end ($y_{t=0} = H)$
of the sample through a rigid sleeper, unless differently specified.
The sides $r = R$ and $y = 0$ represent the wall (container), they
react to the grains actions as described beforehand. However, at $r
= 0$, a symmetry condition is simulated by omitting the friction
effect (Figure \ref{BC}-(a)). Henceforth, this loading case will be
termed as partially confined configuration. The fully confined
configuration corresponds to the case where the excitation is
applied on the whole upper end of the sample.

Under repeated loading granular materials undergo large
displacements towards free regions. It is interesting to notice that
observations of displacement fields and strains showed that granular
materials deform because of rearrangements of the packing, rather
than contact elasticity \cite{Roux, Combe}. Unlike fully confined
samples, where the settlement is mainly due to grain rearrangement
and overwork, the mobility of grains in partially confined granular
materials is predominant. An experimental work performed by the
authors \cite{Karrech, karrech2} on irregular ballasted samples
showed that the acceleration of the sleeper plays a key role on the
mobility of grains. In both confinement cases, it has been shown
that the settlement increases with the acceleration. In addition, in
case of partially confined samples, a sharp increase in terms of
settlement speed has been noticed when the acceleration exceeds the
gravity. At high level of agitation, the loss of contacts is
frequent especially at the critical plane relating the edge of the
sleeper to the wall as can be seen in figure (\ref{BC}-b).

\subsection{Settlement in terms of acceleration and loss of contacts}
\begin{figure}[h!]
\begin{center}
\begin{tabular}{c}
  \includegraphics[width=1\hsize]{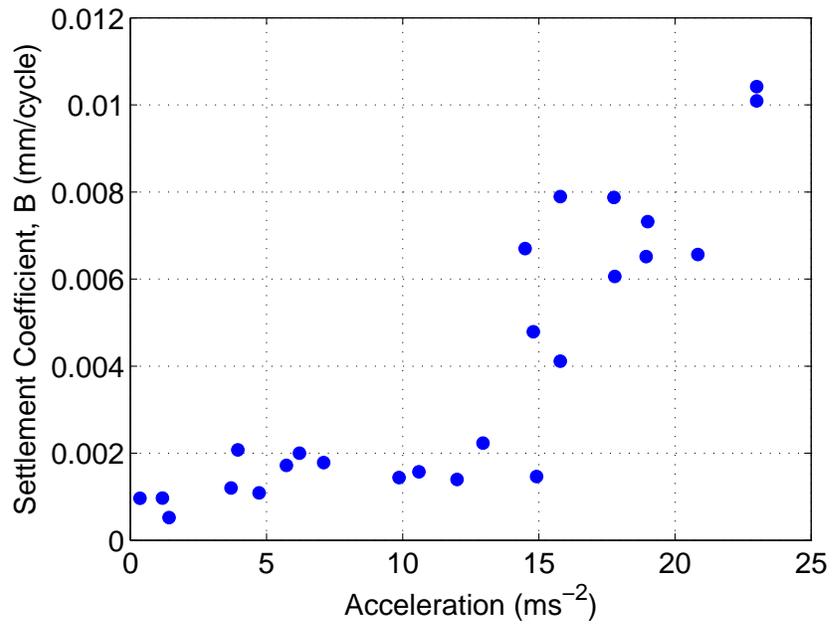}
\end{tabular}
\caption{\label{ouv_B_Acc} Settlement speed with respect to
acceleration.}
\end{center}
\end{figure}

The simulations conducted herein are performed using polygonal
grains in order to produce a shape which is similar to the
micro-ballast used in the above mentioned study \cite{Karrech,
karrech2}. The obtained numerical results consist of axial
displacement of the sleeper with respect to time. The applied forces
are varied from $0.5$ to $3$ kN and the frequency is varied from
$20$ to $40$ Hz, in such a way that the sleeper covers a wide range
of accelerations around the gravity. The grains radii are
distributed uniformly from $r_{min}= 5$ mm to $r_{max} =10$ mm. In
accordance with the experimental procedure, the settlement under
sleeper, $h_{max}$ is described with a logarithmic law of the form:
$h_{max}(N)=A+B \ln (N)$, with respect to the number of cycles $N$.
This logarithmic law seems to be valid for different granular
materials under cyclic loading
\cite{Barksdale,Nowak,Ribiere,shenton} The parameters $A$ and $B$
depend on different physical factors such as the degree of
confinement, the applied force, and the frequency. These factors are
independent and they can affect individually the settlement speed.
The experimental result revealed that the acceleration which depends
simultaneously on the applied force and frequency is the best
explicative quantity in terms of correlation. Moreover, it has been
shown that the degree of confinement results in different behaviors
in terms of settlement versus acceleration. The simulations
conducted herein produce most of the experimental features. It can
be seen that the partially confined configurations exhibit a
critical transition in terms of settlement speed at around $1.4 g$,
as can be seen in figure (\ref{ouv_B_Acc}). This transition has been
observed experimentally \cite{Karrech}, it is probably due to the
loss of contacts which can not be detected easily with the
considered experimental setup. Numerical simulation can provide some
more information regarding the history of the granular material
texture.

\begin{figure}[h!]
\begin{center}
\begin{tabular}{c}
  \includegraphics[width=0.9\hsize]{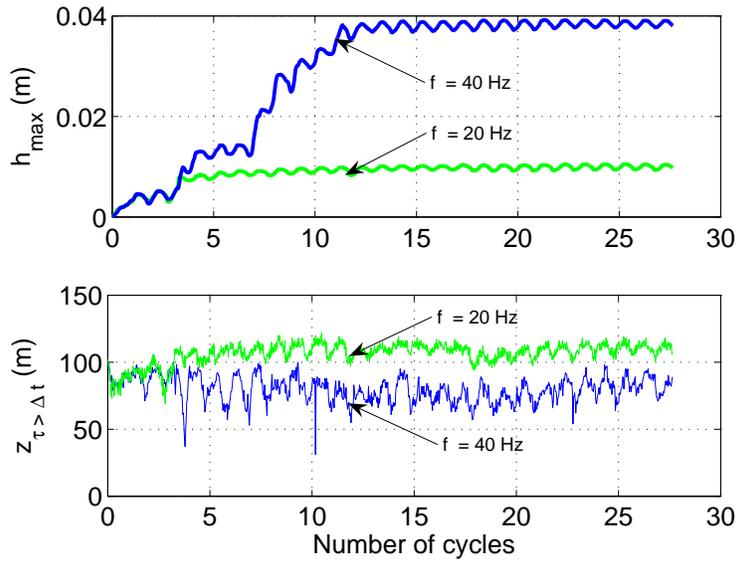}
\end{tabular}
\caption{\label{Acc} Response of the sample in terms of settlement
($h_{max}$) and number of contacts ($Z_{\tau > \Delta t}$) with
respect to the number of cycles at different frequencies.}
\end{center}
\end{figure}

Figure (\ref{Acc}) shows the variation of the sleeper displacement
as well as the number of contacts with respect to the number of
loading cycles, at different frequencies. In these particular cases,
the number of cycles is limited to around $25$, for clarity, and the
applied force amplitude is equivalent to $1.5$ kN. It can be noticed
that the settlement is much higher when the acceleration is beyond
the gravity level ($f=40$ Hz). On the other hand, it can be seen
that the number of contacts follows the exciting force, in terms of
oscillations. This means that while cyclically loaded, the granular
material exhibits local alternating opening and closure of contacts
independently of the acceleration level. However, it can be seen
that there is a difference of about $20 \%$ in terms of permanent
loss of contacts when the sleeper undergoes high acceleration.

\begin{figure}[h!]
\begin{center}
\begin{tabular}{c}
  \includegraphics[width=0.9\hsize]{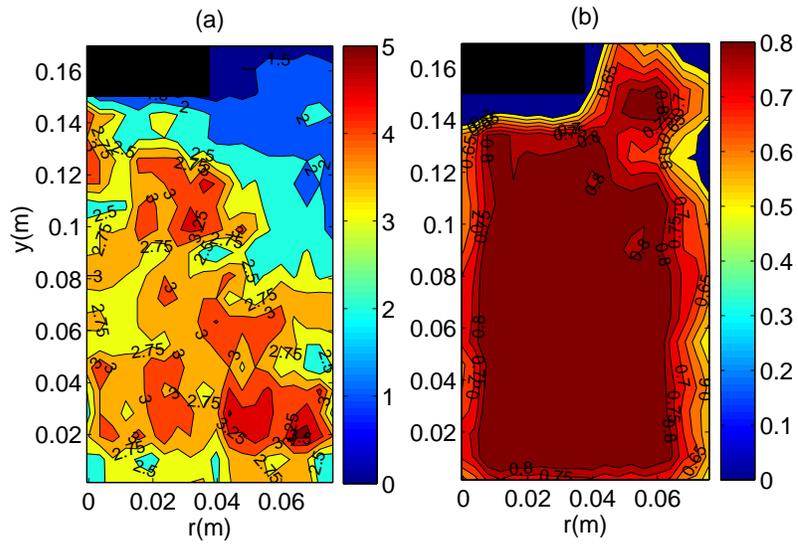}
\end{tabular}
\caption{\label{sample_contact_density} Distribution of (a)
coordination number (b) material density.}
\end{center}
\end{figure}

Under cyclic loading of partially confined sample, the grains flow
towards less loaded regions as can be seen in figure (\ref{BC}-b).
More interestingly, it can be seen in figure
(\ref{sample_contact_density}-a) that during the settlement process,
contact openings occur at a specific region between the front of the
sleeper and the wall frontier. In the same region there is a loss of
density as can be seen in figure (\ref{sample_contact_density}-b).

This critical and localized behavior can be observed during the
settlement process independently of the number of cycles. In the
following section, it will be shown that there is a causality
relationship between the loss of contacts and the settlement speed.

\section{Causality relationship between contacts loss and settlement
speed} In order to find out a logical relationship between the
responses of granular materials under vibration, statistical
analysis are necessary since the solution is not analytically
determinist. In this section, the objective is to investigate the
causality between the settlement velocity and the loss of contacts.
Therefore, we adopt Granger's approach \cite{granger} which was
introduced in econometric in order to forecast possible relationship
between discrete time series. However, this approach can be applied
for physical systems providing accessible responses with respect to
time \cite{chen}. The model questions whether the prediction of a
given variable is improved by taking into account its own history
and the history of another variable. The original model concerns
only stationary time signals (time series where the first and second
moments are independent of time) or linear (time series with a
stationary rate of change). More recently, it has been shown
\cite{Dufour} that it is possible to conduct causality testing on
non stationary time series if (i) they can be approximated by
tendency functions and (ii) the estimation errors are stationary.

\subsection{Causality measurement}
\label{mesure_de_caus} 
\begin{figure}[h!]
\begin{center}
\begin{tabular}{c}
  \includegraphics[width=1\hsize]{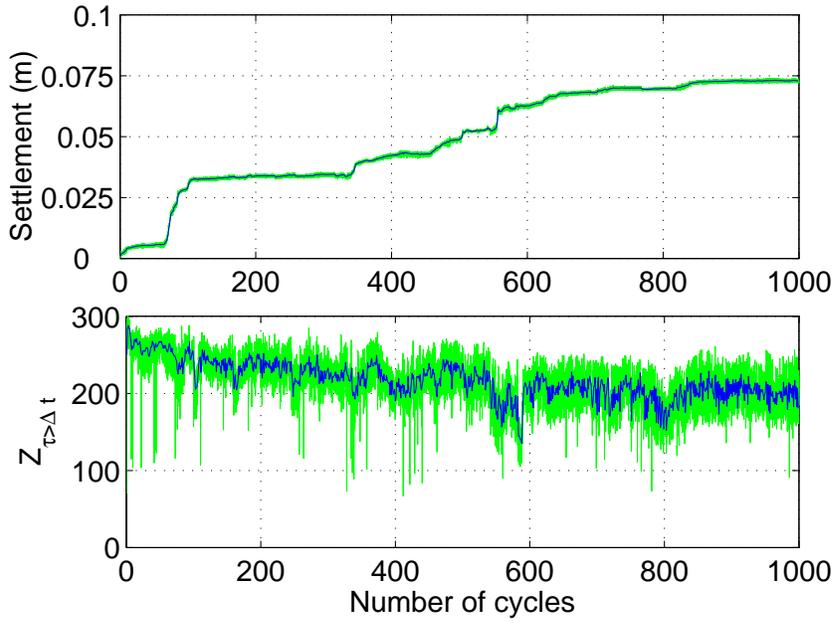}
\end{tabular}
\caption{\label{filtre} Variation of the settlement and of the
number of permanent contacts with respect to the number of cycles.}
\end{center}
\end{figure}

The physical quantities of interest are the
settlement under sleeper, h(t), and the number of permanent
contacts, z(t). In order to illustrate the causality analysis, a
granular sample consisting of polygonal grains of diorite with an
average size of $5$ mm is considered. The applied force in this
particular case is of frequency $60$ Hz and amplitude $1kN$. The
time series are extracted from the numerical results by averaging
over the loading cycles as follows: $z (n) = <z(t)> =
\frac{1}{T}\int\limits_{(n - 1)T}^{nT} {z(t)dt} $ and $h(n)=<h(t)>$,
where T is the loading period (Figure \ref{filtre}). The dynamic
relationship between the above mentioned variables are then
described with a multivariate autoregressive model of the order $l$
as follows:
\begin{equation}
  \label{autoreg_1}
    \begin{array}{l}
 \displaystyle  h(i) = h^r(i) + \sum\limits_{k = 1}^l {a_k h(i - k)}  + \sum\limits_{k = 1}^l {b_k z(i - k)}  + \epsilon _h (i) \\
 \displaystyle  z(i) = z^r(i) + \sum\limits_{k = 1}^l {c_k h(i - k)}  + \sum\limits_{k = 1}^l {d_k z(i - k)}  + \epsilon _z (i) \\
    \end{array}
\end{equation}
where $h^r(i) = \alpha + \beta Ln(i)$ and $z^r(i)=\chi + \gamma i$
are the tendency functions with respect to the number of cycles, and
$\epsilon$ are the error terms, which are assumed to be independent.

In accordance with the Granger's method, a comparaison in terms of
accuracy is conducted between the above description
(\ref{autoreg_1}) and the following model:

\begin{equation}
  \label{autoreg_2}
\begin{array}{l}
\displaystyle  h(i) = \hat h^r(i) + \sum\limits_{k = 1}^l {\hat a_k h(i - k)}  + e_h (i) \\
\displaystyle  z(i) = \hat z^r(i) + \sum\limits_{k = 1}^l {\hat d_k z(i - k)}  + e_z (i) \\
 \end{array}
\end{equation}
where  $\hat h^r(i) = \hat \alpha  + \hat \beta Ln(i)$ and $\hat
z^r(i) = \hat \chi + \hat \gamma i$ are tendency functions with
respect to the number of cycles, and $e$ are the error terms which
are assumed to be independent. The main idea is to compare the
accuracies of the two models. If the description (\ref{autoreg_1})
improves the estimation of the physical quantity $h$ as compared to
the description (\ref{autoreg_2}), that means that $z$ causes $h$,
since $z$ is an explicative variable of $h$. The non causality
hypothesis is expressed by $H_0:$ $b_1 = b_2 = ... = b_l = 0$.
Henceforth, the term ``$H_0$'' will be used to refer to the above
mentioned hypothesis. When this hypothesis is valid, the first term
of the system (\ref{autoreg_1}) is reduced to the first term of
(\ref{autoreg_2}). Therefore, it is possible to calculate the
causality measure of $z$ and $h$ based on the above mentioned
autoregressive models as follows:
\begin{equation}
  \label{autoreg_3}
\begin{array}{l}
\mathcal{F} = (N - 2l - 2) Ln \left(
\frac{\Sigma_{e_s}}{\Sigma_{\epsilon_s}} \right)
 \end{array}
\end{equation}
where N denotes the length of the times series, $\Sigma _{e_s }$ and
$\Sigma _{\epsilon _s }$ represent the auto-variances of $e_s$ and
$\epsilon _s$ where $s = h,z$. Under the nullity hypothesis $H_0$,
the measure $\mathcal{F}_{z \to h}$ has an asymptotic $\chi ^2(2l)$
distribution \cite{crei}. In order to reject the hypothesis $H_0$
with a preselected risk level $\alpha$, the calculated $\chi^2$
should be higher than the value of the standard distribution of
parameters $\alpha$ and $l$).

\subsection{Estimation coefficients}
\begin{figure}[h!]
\begin{center}
\begin{tabular}{c}
  \includegraphics[width=1\hsize]{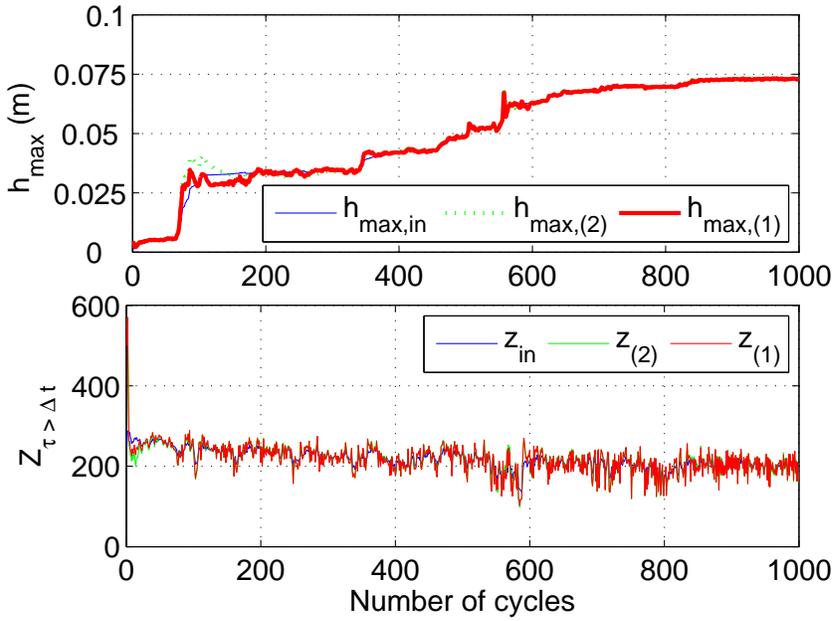}
\end{tabular}
\caption{\label{causality} Application of the models
(\ref{autoreg_1}) and (\ref{autoreg_2}) on the signals of settlement
and permanent contacts. The subscripts ``in'', ``1'', and ``2''
denote the MD method  calculations, the estimation using the first
model and the estimation using the second model, respectively.}
\end{center}
\end{figure}

The causality measure between the considered signals can be
calculated using the error terms $\epsilon_s$ and $e_s$, where $s =
h,z$. Therefore, it is necessary to estimate first of all the
coefficients of the suggested models and deduce the error terms.
This task can be accomplished using the recursive least square
method \cite{beex}. In this section, we adapt our description to the
algorithm by expressing the signals under consideration as follows:
\begin{equation}
    \label{autoreg_4}
    \begin{array}{l}
\displaystyle d(i) = \boldsymbol{A}^{t} \boldsymbol{x}^{(i)} + u(i)
    \end{array}
\end{equation}
where $d(i)$ is an output corresponding to $h$, $z$ or both of them
as will be seen later on, $\boldsymbol{A}$ is a vector which
encloses the unknown estimation parameters and $u(i)$ represents the
estimation error at the step $i$. In the case of model
(\ref{autoreg_1}), these vectors can be written as an input
$\boldsymbol{x}^{(i)} = [ 1, Ln(i) ,h(i - 1),.., h(i - l),1 ,i, z(i
- 1),..,z(i - l)]$ , an unknown $\boldsymbol{A} = [\alpha, \beta
,a_1,.., a_l,\chi ,\gamma, b_1,.., \allowbreak b_l]$ and error terms
$u(i)=\epsilon_h(i)$. Similarly, in the case of model
(\ref{autoreg_2}), these vectors can be written as
$\boldsymbol{x}^{(i)} = [1, Ln(i) ,h(i - 1),.., h(i - l)]$,
$\boldsymbol{A} = [\alpha, \beta ,a_1,.., a_l]$ and $u(i)=e_h(i)$.
In order to estimate the unknown parameters of the model
(\ref{autoreg_4}), the least squares criterion is used:
\begin{equation}
    \label{autoreg_5}
    \begin{array}{l}
\displaystyle     \mathcal{R} = \sum_{i=0}^{n} \lambda^{n-i} \left(
u(i) \right)^2
    \end{array}
\end{equation}
Solving the problem consists in minimizing the quantity
$\mathcal{R}$ with respect to the unknown vector $\boldsymbol{A}$.
This leads to a recursive algorithm (appendix), which consists in
starting from an initial set of parameters $\boldsymbol{A}
=\boldsymbol{0} $, an initial matrix $\boldsymbol{Q} =
\boldsymbol{I}$ of size $l \times l$ and a real constant $0 <<
\lambda <1$. At each iteration $n$, the calculation steps read:
\begin{equation}
    \label{autoreg_6}
    \begin{array}{l}
        \boldsymbol{k} = \frac{\lambda^{-1} \boldsymbol{Q} \boldsymbol{x}}{1+ \lambda^{-1} \boldsymbol{x} \boldsymbol{Q}
        \boldsymbol{x}^t}\\
        u(n) = d(n)- \boldsymbol{x}^t \boldsymbol{A}\\
        \boldsymbol{A}=\boldsymbol{A}+\boldsymbol{k} u(n)\\
        \boldsymbol{Q} =\lambda^{-1} \left( \boldsymbol{Q} - \boldsymbol{k} \boldsymbol{x}^t \boldsymbol{Q}
        \right)\\
    \end{array}
\end{equation}
where the vector $\boldsymbol{k}$ is termed as the Kalman gain.
Knowing the parameters of the models (\ref{autoreg_1}) and
(\ref{autoreg_2}), it is possible to plot out the estimations of the
settlement and permanent contacts number as shown in figure
(\ref{causality}).

Once the error terms are obtained, the causality measure can be
calculated using equation (\ref{autoreg_3}). In the particular case
considered in this study, the discrete time signals $h$ and $z$ are
of length $N=1000$ cycles and the order of autoregression is of
$l=N/100$. The risk level of rejecting the hypothesis $H_0$ is
$\alpha=1\%$. At this risk level, and for a number of parameters
$l$, the calculated theoretical value of $\chi^2(2l)$ is $37.52$,
however, the measure of causality equals $590.85$. Therefore, it can
be concluded that the number of permanent contacts loss is a
significant explication variable of the settlement under sleeper.
This approach can be applied for different loading cases. Table
(\ref{tab_caus}) shows that the causality direction remains valid
for different frequencies and amplitudes of loading.

\section{Conclusion}
In this study, numerical simulations of a partially confined
granular material under vibration are presented. It has been shown
that the settlement process is characterized by a flow of materials
towards less loaded regions. It has also been noticed that there is
privileged regions of contacts loss. Furthermore, it has been proved
that the loss of contacts causes the settlement, for different
loading cases. The residual displacements which take place under
dynamic loading at different frequencies and amplitudes depend on
several factors such as material properties, grain shape, degree of
confinement etc. In this study, we selected the diorite and the
polygonal shape because this type of material is widely used in
railway platforms. Moreover, we concentrated on the loss of contacts
as an explicative variable of settlement. As a perspective of the
suggested analysis, it would be interesting to investigate all the
physical factors which may influence the settlement speed, such as
the acceleration and elastic deflection. It is also possible to
extend the suggested procedure for more complex grain shapes. For
instance, as long as the settlement depends on the mobility of the
material, non convex grains may have an important effect on the
settlement. Actually, unlike convex grains (circular, polygonal,
elliptic etc.) where the contact are binary, non convex grains can
be connected with more than a single contact. Assuming equivalent
sizes of contact areas, the dissipation should be higher in case of
non convex grain. In addition, the contact openings are more
difficult to take place, therefore the settlement should be
different. This particular point will be the subject of a future
contribution by the authors.

\appendix
\section{Recursive Least Squares} Minimizing the least square criterion (\ref{autoreg_5}) leads to
the following equation:
\begin{equation}
    \begin{array}{l}
\displaystyle     \sum_{i=0}^{n} \lambda^{n-i} x(i)
\boldsymbol{x}^{(i)} = \sum_{i=0}^{n} \lambda^{n-i} \left[
    \boldsymbol{x}^{(i)} \boldsymbol{x}^{(i)^t}\right]\boldsymbol{A}^{(n)}
    \end{array}
\end{equation}
Let $p^{(n)}=\sum_{i=0}^{n} \lambda^{n-i} x(i)
\boldsymbol{x}^{(i)}$, $R^{(n)}=\sum_{i=0}^{n} \lambda^{n-i}
\big[\boldsymbol{x}^{(i)} \boldsymbol{x}^{(i)^t} \big]$. Using the
Woodbury identity of matrices (Kima et Bennighof \cite{woodbury})
one can obtain the following relationship:
\begin{equation}
    \label{autoreg_7}
    \begin{array}{l}
\displaystyle \boldsymbol{Q}^{(n)} = \lambda^{-1}
\boldsymbol{Q}^{(n-1)}-\lambda^{-1} \boldsymbol{k}^{(n)}
\boldsymbol{x}^{(n)^t} \boldsymbol{Q}^{(n-1)}
    \end{array}
\end{equation}
where $\boldsymbol{Q}^{(n)}=\boldsymbol{R^{-1}}$$^{(n)}$ and
$\boldsymbol{k}^{(n)}=\frac{\lambda^{-1} \boldsymbol{Q}^{(n-1)
\boldsymbol{x}^{(n)}}}{1+\lambda^{-1} \boldsymbol{x}^{(n)^t}
\boldsymbol{Q}^{(n-1)} \boldsymbol{x}^{(n)}}$. Introducing this
expression in the equation $\boldsymbol{A}^{(n)} =
\boldsymbol{Q}^{(n)} \boldsymbol{p}^{(n)}$, leads to a solution of
the problem in a recurrent formula:
\begin{equation}
    \label{autoreg_8}
    \begin{array}{l}
\displaystyle \boldsymbol{h}^{(n)} = \boldsymbol{h}^{(n-1)} +
\boldsymbol{k}^{(n)} u(n)
    \end{array}
\end{equation}

\bibliographystyle{spbasic}
\bibliography{main}   

\begin{thebibliography}{26}
\providecommand{\natexlab}[1]{#1}
\providecommand{\url}[1]{{#1}}
\providecommand{\urlprefix}{URL }
\expandafter\ifx\csname urlstyle\endcsname\relax
  \providecommand{\doi}[1]{DOI~\discretionary{}{}{}#1}\else
  \providecommand{\doi}{DOI~\discretionary{}{}{}\begingroup
  \urlstyle{rm}\Url}\fi
\providecommand{\eprint}[2][]{\url{#2}}

\bibitem[{Allen and Tildesley(1989)}]{allen}
Allen M, Tildesley D (1989) Computer Simulation of Liquids. Oxford University
  Press, Bristol

\bibitem[{Azema et~al(2006)Azema, Radjai, Peyroux, Dubois, and
  Saussine}]{azema}
Azema E, Radjai F, Peyroux R, Dubois F, Saussine G (2006) Vibrational dynamics
  of confined granular material. Physical Review E 74:031,302--031,312

\bibitem[{Barksdale(1972)}]{Barksdale}
Barksdale RD (1972) Laboratory evaluation of rutting in base course materials.
  Third International Conference on Structural Design of Asphalt Pavement,
  September 1972, London, England 3:161--174

\bibitem[{Beex and Zeidler(2003)}]{beex}
Beex AA, Zeidler JR (2003) Non-wiener effects in recursive least squares
  adaptation. IEEE pp 7803--7946

\bibitem[{Chen et~al(2004)Chen, Rangarajan, Feng, and Ding}]{chen}
Chen Y, Rangarajan G, Feng J, Ding M (2004) Analyzing multiple nonlinear time
  series with extended granger causality. Physics Lettrs A 324:26--35

\bibitem[{Ciamarra et~al(2005)Ciamarra, Coniglio, and Nicodemi}]{SegMix}
Ciamarra MP, Coniglio A, Nicodemi M (2005) Shear-induced segregation of a
  granular mixture under horizontal oscillation. Journal of Physics Condensed
  Matter 17(24):2549--2556

\bibitem[{Clément et~al(1995)Clément, Rajchenbach, and Duran}]{Mixing}
Clément E, Rajchenbach J, Duran J (1995) Mixing of a granular material in a
  bidimensional rotating drum. Materials Research Society Symposium -
  Proceedings 367(13):513--517

\bibitem[{Combe and Roux(2000)}]{Combe}
Combe G, Roux JN (2000) Strain versus stress in a model granular material: A
  devil's staircase. Physical Review Letters 85(17):3628--3631

\bibitem[{Cundall and Strack(1979)}]{Cundall}
Cundall PA, Strack ODL (1979) A discrete numerical model for granular
  assemblies. Géotechnique 29:47--65

\bibitem[{Dufour et~al(2007)Dufour, Pelletier, and Renault}]{Dufour}
Dufour J, Pelletier D, Renault E (2007) Multivariate out-of-sample tests for
  granger causality. Computational Statistics and Data Analysis 51:3319--3329

\bibitem[{Gandolfo and Nicolettiy(2002)}]{crei}
Gandolfo G, Nicolettiy G (2002) Exchange rate volatility and economic openness:
  A causal relation ? Centro ricerche di economia internazionale 68

\bibitem[{Granger(1969)}]{granger}
Granger C (1969) Investigating causal relations by econometric models and cross
  -spectral methods. Econometrica 37:424--459

\bibitem[{Karrech(2007)}]{Karrech}
Karrech A (2007) Comportement des matériaux granulaires sous vibration:
  Application au cas du ballast. PhD thesis, Ecole Nationale des Ponts et
  Chaussées, France

\bibitem[{Karrech et~al(2006{\natexlab{a}})Karrech, Duhamel, Bonnet, Canou,
  Dupla, Roux, and Chevoir}]{karrech2}
Karrech A, Duhamel D, Bonnet G, Canou J, Dupla JC, Roux JN, Chevoir F
  (2006{\natexlab{a}}) Experimental study of settlement mechanisms in
  micro-ballast beds under dynamic loading. Submitted to the Journal of
  Materials in Civil Engineering

\bibitem[{Karrech et~al(2006{\natexlab{b}})Karrech, Duhamel, Bonnet, Roux,
  Chevoir, Sab, Canou, and Dupla}]{Karrech3}
Karrech A, Duhamel D, Bonnet G, Roux JN, Chevoir F, Sab K, Canou J, Dupla JC
  (2006{\natexlab{b}}) Discrete element method for granular materials
  preparation and settlement. The Third International Conference on Advances In
  Mechanical Engineering and Mechanics, Hammamet, Tunisia, December 17-19

\bibitem[{Karrech et~al(2007)Karrech, Duhamel, Bonnet, Roux, Chevoir, Canou,
  Dupla, and Sab}]{Karrech1}
Karrech A, Duhamel D, Bonnet G, Roux JN, Chevoir F, Canou J, Dupla JC, Sab K
  (2007) A computational procedure for the prediction of settlement in granular
  materials under cyclic loading. Computer Methods in Applied Mechanics and
  Engineering 197:80--94

\bibitem[{Kima and Bennighof(2006)}]{woodbury}
Kima CW, Bennighof J (2006) Fast frequency response analysis of partially
  damped structures with non-proportional viscous damping. Journal of Sound and
  Vibration 297:1075--1081

\bibitem[{Lobo-Guerrero and Vallejo(2006)}]{lobo_vall}
Lobo-Guerrero S, Vallejo L (2006) Discrete element method analysis of railtrack
  ballast degradation. Granular Matter 8:195--204

\bibitem[{Lu and McDowell(2007)}]{Lu_Mc}
Lu M, McDowell GR (2007) The importance of modelling ballast particle shape in
  the discrete element method. Granular Matter 9:69--80

\bibitem[{Moon et~al(2003)Moon, Goldman, Swinney, and Swift}]{Transport}
Moon SJ, Goldman DI, Swinney HL, Swift JB (2003) Kink-induced transport and
  segregation in oscillated granular layers. Physical Review Letters
  91(13):1343,011--1343,014

\bibitem[{Nowak et~al(1998)Nowak, Knight, Ben-Naim, Jaeger, and Nagel}]{Nowak}
Nowak ER, Knight JB, Ben-Naim E, Jaeger HM, Nagel SR (1998) Density
  fluctuations in vibrated granular materials. Physical Review E Statistical
  Physics 57(2):1971--1982

\bibitem[{Ribiere et~al(2005)Ribiere, Richard, Bideau, and Delannay}]{Ribiere}
Ribiere P, Richard P, Bideau D, Delannay R (2005) Experimental compaction of
  anisotropic granular media. The European Physical Journal E 16:415--420

\bibitem[{Rosato and Doris(2000)}]{Rosatocompaction}
Rosato AD, Doris Y (2000) Microstructure evolution in compacted granular beds.
  Powder Technology 109:255--261

\bibitem[{Roux and Chevoir(2005)}]{Roux}
Roux JN, Chevoir F (2005) Simulation numérique discrète et comportement
  mécanique des matériaux granulaires. Bulletin des Laboratoires des Ponts et
  Chaussées (254):109--138

\bibitem[{Saussine et~al(2005)Saussine, Cholet, Gautier, Dubois, Bohatier, and
  Moreau}]{saussine}
Saussine G, Cholet C, Gautier P, Dubois F, Bohatier C, Moreau J (2005)
  Modelling ballast behavior under dynamic loading, part1: A 2d polygonal
  discrete element method approach. Computer Method in Applied Mechanics and
  Enginnering 195(19-22):2841--2859

\bibitem[{Shenton(1978)}]{shenton}
Shenton M (1978) Deformation of Railway Ballast under Repeated Loading
  Conditions. Oxford

\end{thebibliography}

\newpage
\listoftables

\newpage

\listoffigures

\newpage

\begin{table}[hbtp]
\begin{center}
\begin{tabular}{l c c}
\hline \hline Frequency, $f$ (Hz)& Applied Force, $\Delta F$ (kN)& $\chi^2$ \\
\hline
   & 0.5 & 621.57 \\
20 & 1 & 227.61 \\
   & 1.5 & 457.06 \\
   & 0.5 & 634.34 \\
30 & 1 & 750.91 \\
   & 1.5 & 640.1 \\
   & 0.5 & 703.2\\
40 & 1 & 430.9\\
   & 1.5 & 685.42\\

\hline
\end{tabular}
\caption{ Causality measure for different excitations}
\label{tab_caus}
\end{center}
\end{table}

\newpage

\newpage

\newpage

\newpage

\newpage

\newpage
\newpage

\end{document}